\documentclass[journal=jacsat,manuscript=article]{achemso}

\usepackage[version=3]{mhchem} 
\usepackage{xcolor}
\usepackage{amsmath}

\author{Tao He}
\affiliation[East China Normal University]{State Key Laboratory of Precision Spectroscopy Science and Technology, East China Normal University, Shanghai 200241, China.}

\author{Haoran Liu}
\affiliation[East China Normal University]{State Key Laboratory of Precision Spectroscopy Science and Technology, East China Normal University, Shanghai 200241, China.}

\author{Zihe Jiang}
\affiliation[East China Normal University]{State Key Laboratory of Precision Spectroscopy Science and Technology, East China Normal University, Shanghai 200241, China.}

\author{Zhiwei Hu}
\affiliation[East China Normal University]{State Key Laboratory of Precision Spectroscopy Science and Technology, East China Normal University, Shanghai 200241, China.}

\author{Banghuan Zhang}
\affiliation[East China Normal University]{State Key Laboratory of Precision Spectroscopy Science and Technology, East China Normal University, Shanghai 200241, China.}

\author{Xiaohui Dong}
\affiliation[East China Normal University]{State Key Laboratory of Precision Spectroscopy Science and Technology, East China Normal University, Shanghai 200241, China.}

\author{Chaowei Sun}
\affiliation[Henan Academy of Sciences]{Institute of laser Manufacturing, Henan Academy of Sciences, Zhengzhou 450046, China.}

\author{Wei Jiang}
\affiliation[Henan Academy of Sciences]{Institute of laser Manufacturing, Henan Academy of Sciences, Zhengzhou 450046, China.}

\author{Jiawei Sun}
\affiliation[Shenzhen University]{State Key Laboratory of Radio Frequency Heterogeneous Integration, College of Electronics and Information Engineering, Shenzhen University, Shenzhen 518060, China.}

\author{Yang Li}
\affiliation[Shenzhen University]{State Key Laboratory of Radio Frequency Heterogeneous Integration, College of Electronics and Information Engineering, Shenzhen University, Shenzhen 518060, China.}

\author{Huatian Hu}
\email{huatian.hu@iit.it}
\affiliation[Istituto Italiano di Tecnologia]{Center for Biomolecular Nanotechnologies, Istituto Italiano di Tecnologia, via Barsanti 14, Arnesano, 73010, Italy.}

\author{Wen Chen}
\email{wchen@lps.ecnu.edu.cn}
\affiliation[East China Normal University]{State Key Laboratory of Precision Spectroscopy Science and Technology, East China Normal University, Shanghai 200241, China.}

\author{Hongxing Xu}
\email{hxxu@hnas.ac.cn}
\affiliation[Henan Academy of Sciences]{Institute of laser Manufacturing, Henan Academy of Sciences, Zhengzhou 450046, China.}

\title[An \textsf{achemso} demo]
  {Quantitative Benchmarking of Remote Excitation in Plasmonic Sensing with Enhanced Signal-to-Noise Ratio}

\abbreviations{IR,NMR,UV}
\keywords{American Chemical Society, \LaTeX}

\begin{document}

\begin{tocentry}

Some journals require a graphical entry for the Table of Contents.
This should be laid out ``print ready'' so that the sizing of the
text is correct.

Inside the \texttt{tocentry} environment, the font used is Helvetica
8\,pt, as required by \emph{Journal of the American Chemical
Society}.

The surrounding frame is 9\,cm by 3.5\,cm, which is the maximum
permitted for  \emph{Journal of the American Chemical Society}
graphical table of content entries. The box will not resize if the
content is too big: instead it will overflow the edge of the box.

This box and the associated title will always be printed on a
separate page at the end of the document.

\end{tocentry}

\begin{abstract}
Remote excitation using guided optical modes—such as waveguides, fibers, or surface waves—offers a promising alternative to direct optical excitation for surface-enhanced Raman scattering (SERS), particularly in applications requiring reduced heating, minimal invasiveness, and on-chip integration. However, despite its widespread use, systematic comparisons between remote and direct excitation remain limited. Here, we quantitatively benchmark both schemes by measuring power-dependent SERS responses from individual plasmonic nanogaps. We statistically analyze the maximum achievable SERS intensity before structural degradation, extract local temperatures, and evaluate signal-to-noise ratios (SNR). Our findings reveal that both remote and direct SERS share a common electric-field limit, despite exhibiting different levels of heating. This
suggests that spectral evolution is primarily governed by the local electric field, which drives nanoscale atomic migration rather than excessive heating. Nonetheless, the lower heating associated with remote excitation enhances the Raman SNR by approximately 30\%, improving measurement quality without compromising signal strength. This study establishes a quantitative framework for evaluating excitation strategies in plasmonic sensing, and challenges common assumptions about the role of heating in nanostructural stability under strong optical excitation.
\end{abstract}

\section{Introduction}
Spatially separated (nonlocal) optical sensing, in contrast to its conventional \textit{in situ} counterpart, utilizes guided optical modes to remotely excite and probe the analyte via light-matter interactions. 
This strategy has gained increasing attention due to its less invasive nature \cite{yan_nanowire-based_2012,lu_live-cell_2014,huang_nanowire-supported_2014,zhang_propagating_2016,pisano_vibrational_2025}, improved compatibility with integrated photonic platforms~\cite{redolat_polarization-insensitive_2025,measor_-chip_2007,peyskens_surface_2016}, and ability to disentangle signals that are otherwise mixed in local measurements~\cite{su_visualization_2015,lei_deep_2024,coca-lopez_remote_2018,huang_nanowire-supported_2014,niu_unified_2022}. 
By routing optical fields through fibers~\cite{bello_fiber-optic_1990,chen_gold_2014,zheng_toward_2023,zheng_tunable_2024,pisano_vibrational_2025}, optical and polaritonic waveguides~\cite{zhang_propagating_2016,huang_nanowire-supported_2014,zhang_coherent_2020,huang_remote_2011}, guided-mode excitation mitigates unwanted effects of direct exposure--such as local heating, and enables access to environments that are physically obstructed or sensitive to direct illumination, such as deep inside biological tissues and cells~\cite{pisano_vibrational_2025,zheng_toward_2023,yan_nanowire-based_2012,lu_live-cell_2014}.
When integrated with tapered structures, optical modes~\cite{chen_gold_2014,zheng_toward_2023,zheng_tunable_2024,li_direct_2014,long_plasmonic_2022} can be confined into tiny modal volumes, bridging the dimensional mismatch between excitation beams and nanoscale analytes to enhance interaction strengths. 

Plasmons, arising from the quantum hybridization between collective electron oscillations and photons, offer a powerful means to confine optical fields beyond the diffraction limit at the nanoscale. 
In particular, metallic nanostructures that support both localized and propagating plasmons offer a natural platform for implementing remote excitation schemes~\cite{zhang_propagating_2016,hu_nanoparticle--mirror_2022}.
In this context, nanogap antennas—such as the nanoparticle-on-mirror (NPoM) \cite{lei_revealing_2012,ciraci_probing_2012}, bowties \cite{kinkhabwala_large_2009}, and nanoparticle-on-slits \cite{chen_continuous-wave_2021,redolat_polarization-insensitive_2025,hu_plasmonic_2025}—generate intense electromagnetic hotspots within their metal–insulator–metal (MIM) nanojunctions, enabling the sensing of ultraweak optical signals. \cite{nam2016plasmonic,baumberg_extreme_2019,liBoostingLightMatterInteractions2024a}
Among these applications, plasmon-enhanced surface-enhanced Raman scattering (SERS), capable of reaching single-molecule sensitivity \cite{xuSpectroscopySingleHemoglobin1999a,nie_probing_1997,lim2010nanogap}, serves as a powerful technique for identifying molecular fingerprints in chemical and biological detection.
Conventionally, SERS excitation is performed via direct, local illumination, which ensures strong near-field enhancement but may induce photothermal effects that risk sample damage.
Recently, spatially-separated nonlocal SERS sensing with a guide-mode excitation \cite{hu2024perspective,liao_quantifying_2023,evans_quantifying_2017,song_remote-excitation_2013,fang_remote-excitation_2009,huang_nanowire-supported_2014,li_duplicating_2020,hu_nanoparticle--mirror_2022,redolat_polarization-insensitive_2025} has attracted significant attention for its potential in on-chip integration and biosensing applications, and reducing the risk of damage.

Despite the increasing adoption of guided-mode excitation in plasmonic sensing, a self-consistent, systematic, and quantitative comparison \cite{liao_quantifying_2023,evans_quantifying_2017} between guided-mode and direct excitation remains relatively insufficient.
Previous reports have presented scattered evidence of increased signal-to-noise ratios (SNR), improved spectral stability, and background signal separation under different excitation schemes~\cite{li_duplicating_2020,evans_quantifying_2017,liao_quantifying_2023,hutchison_subdiffraction_2009}. However, other studies have reported diminished or comparable performance~\cite{hu_nanoparticle--mirror_2022,coca-lopez_remote_2018}.
Several studies have emphasized that using guided-mode excitation can help prevent localized heating and thermal accumulation, thereby protecting analytes and stabilizing system performance~\cite{li_duplicating_2020, evans_quantifying_2017, zhang_propagating_2016, huang_nanowire-supported_2014}. However, other evidence suggests that under intense excitation in plasmonic nanocavities, the atomic-scale metallic structures may be more susceptible to local reconstruction than the molecules themselves~\cite{xomalis_controlling_2020,sigle_monitoring_2015,mertens_light_2017,hu2025alchemically}, and controlling the localized temperature is not sufficient to stabilize the optical performance \cite{benzSinglemoleculeOptomechanicsPicocavities2016a}.
Therefore, it remains under discussion whether the concept of local-heat-suppressed remote excitation can benefit SERS and other nonlinear plasmonic applications, which require both intense electric fields and photostable nanogaps to achieve stronger signals.
These disputed results underscore the need for systematic benchmarking to elucidate the practical advantages and limitations of guided-mode remote excitations.

In this work, we present a statistical benchmarking of remote versus direct continuous-wave (cw) laser excitation schemes using a prototypical and well-defined plasmonic nanogap system—a molecule-functionalized NPoM antenna. Systematic \textit{in situ }power-dependent SERS measurements were performed under both excitation schemes until irreversible structural reconstruction (or damage) occurred, evidenced by anomalies in the SERS signal or plasmonic resonance.  
This approach quantitatively revealed the nanogap's damage thresholds and maximum achievable SERS intensities for each excitation scheme. It compared and correlated the power-dependent SNR and nanogap-localized temperature evolution, measured by molecular and electronic Raman thermometry. The results show that, while remote excitation suppressed photothermal heating, leading to an approximate 30\% improvement in SNR, it did not enhance the ultimate SERS intensity limit---both saturated approximately at the same level.
These counterintuitive findings reveal that photothermal effects are not the primary bottleneck limiting maximum signal intensity or triggering damage. Instead, processes driven by intense local fields—likely inducing atomic-scale structural reconstruction or morphological degradation within the nanogap—dominate the response at high SERS intensities. 
This work establishes a quantitative framework for selecting optimal excitation strategies in plasmonic sensing in emerging remote sensing platforms.

\section{Results and discussion}

\begin{figure}[!h]
  \centering
  \includegraphics[width=1\textwidth]{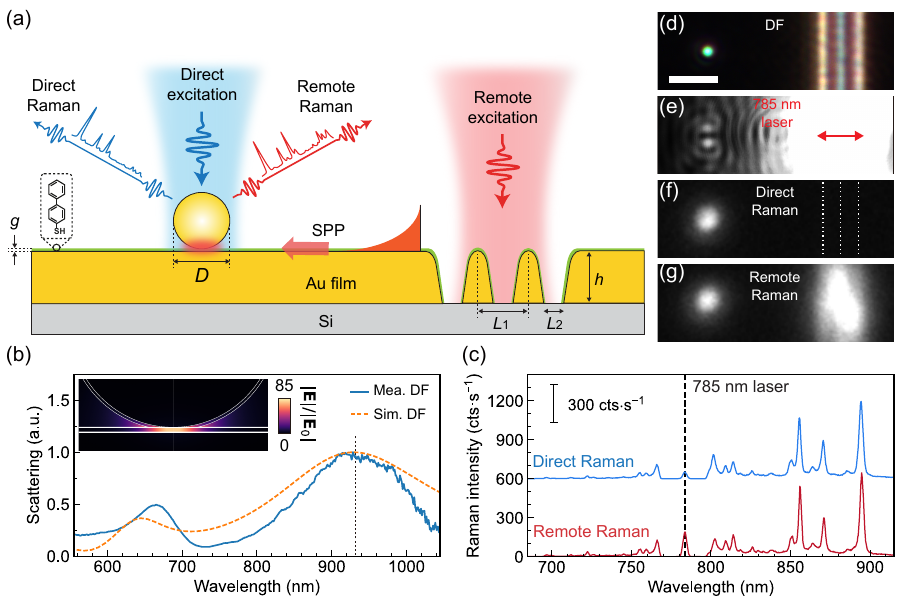}
  \caption{\textbf{Schematics for the remote and local excitation schemes.} (a) Schematic illustration of the remote and direct SERS sensing from a NPoM system with a monolayer of self-assembled BPT molecules as a spacer ($g$ = 1.3 nm). Nanoparticle diameter $D=150$ nm, the sides and height of the nanoslits are $L_1=750$ nm (pitch), $L_2=250$ nm, $h=100$ nm. In the remote excitation scheme, SPPs are launched by illuminating nanoslits. It subsequently propagates several micrometers along the metal surface to excite the NPoM nanogaps. The colors of the laser in the schematic are used only to distinguish excitation schemes and do not represent actual wavelengths. SERS analysis under both schemes was performed with a 785 nm laser. (b) Measured (blue) and simulated (orange) dark-field scattering spectrum of the 150 nm NPoM nanogap. (c) Raman spectra of the 150 nm NPoM under direct (blue) and remote (red) excitation. The blue-line spectrum is offset upward for clear visualization. The black dashed line marks the 785 nm laser wavelength. (d) The dark field image of a 150-nm NPoM structure and the nanoslits. (e) The optical image under the 785-nm laser excitation. The red arrow denotes the laser polarization. (f, g) Raman emission images of the device under direct excitation by illuminating the NPoM structure (f), and under remote excitation by illuminating the nanoslits (g). }
  \label{fig:1}
\end{figure}

Figure~\ref{fig:1}a presents a schematic overview of the sample architecture and optical characterization strategy employed to compare both direct and remote excitation of plasmonic nanogaps. The nanogap is realized using a NPoM configuration \cite{lei_revealing_2012}, where a 150 nm diameter gold nanoparticle is positioned on top of a 100 nm thick Au film, separated by a 1.3 nm thick self-assembled monolayer (SAM) of biphenyl-4-thiol (BPT) molecules (see Supplementary S1 for fabrication details). 
This well-defined vertical nanogap geometry supports hybridized plasmonic gap modes \cite{lei_revealing_2012,tserkezisHybridizationPlasmonicAntenna2015}, manifesting as prominent resonance peaks observed in experimental and simulated scattering spectra (Fig.~\ref{fig:1}b, see Methods for the details).
The resulting localized surface plasmon resonance yields strong electromagnetic local field confinement in the gap, with near-field enhancements reaching up to a factor of $\sim85$, as shown in the inset of Fig.~\ref{fig:1}b. (See details of numerical simulation in Methods.)
In this architecture, BPT molecules simultaneously serve as stable nanometer-scale spacers and SERS probes, enabling real-time optical monitoring of the gap environment.

Under direct excitation, the NPoM acts as an optical nanoantenna: it efficiently captures incident far-field light, concentrates it into an extremely localized field within the nanogap, and subsequently radiates the resulting SERS emission from the confined BPT molecules back into the far field.
Notably, the 150 nm NPoM has a dominant plasmonic resonance spanning from 700 to 1000 nm (Fig. \ref{fig:1}b), enhancing both the Stokes and anti-Stokes Raman bands associated with 785 nm excitation, resulting in high-SNR SERS spectra (Fig.~\ref{fig:1}c) and image (Fig.~\ref{fig:1}f).

In contrast, the remote excitation configuration uses a periodic slit grating patterned on the Au film substrate to enable efficient plasmon launching. This grating couples incident free-space light into high-momentum surface plasmon polaritons (SPPs) at the Au/BPT/air interfaces via momentum matching (see AFM images and fabrication details in Supplementary S1).
Under the 785 nm laser illumination with the polarization orthogonal to slits, the grating launches SPPs that propagate on the Au surface and subsequently couple into the NPoM gap modes \cite{hu_nanoparticle--mirror_2022}, thereby driving the SERS process in an \textit{indirect} manner (Fig.~\ref{fig:1}e). Similarly, the generated SERS signal is then emitted from the NPoM and collected in the far field. 
This remote excitation pathway also produces clear SERS signals (Fig.~\ref{fig:1}c), as visualized by Raman emission image (Fig.~\ref{fig:1}g). Since the NPoMs act as nanoantennas that can be efficiently excited by both schemes, a direct comparison between remote and direct excitation is feasible.



\begin{figure}[!ht]
  \centering
  \includegraphics[width=0.8\textwidth]{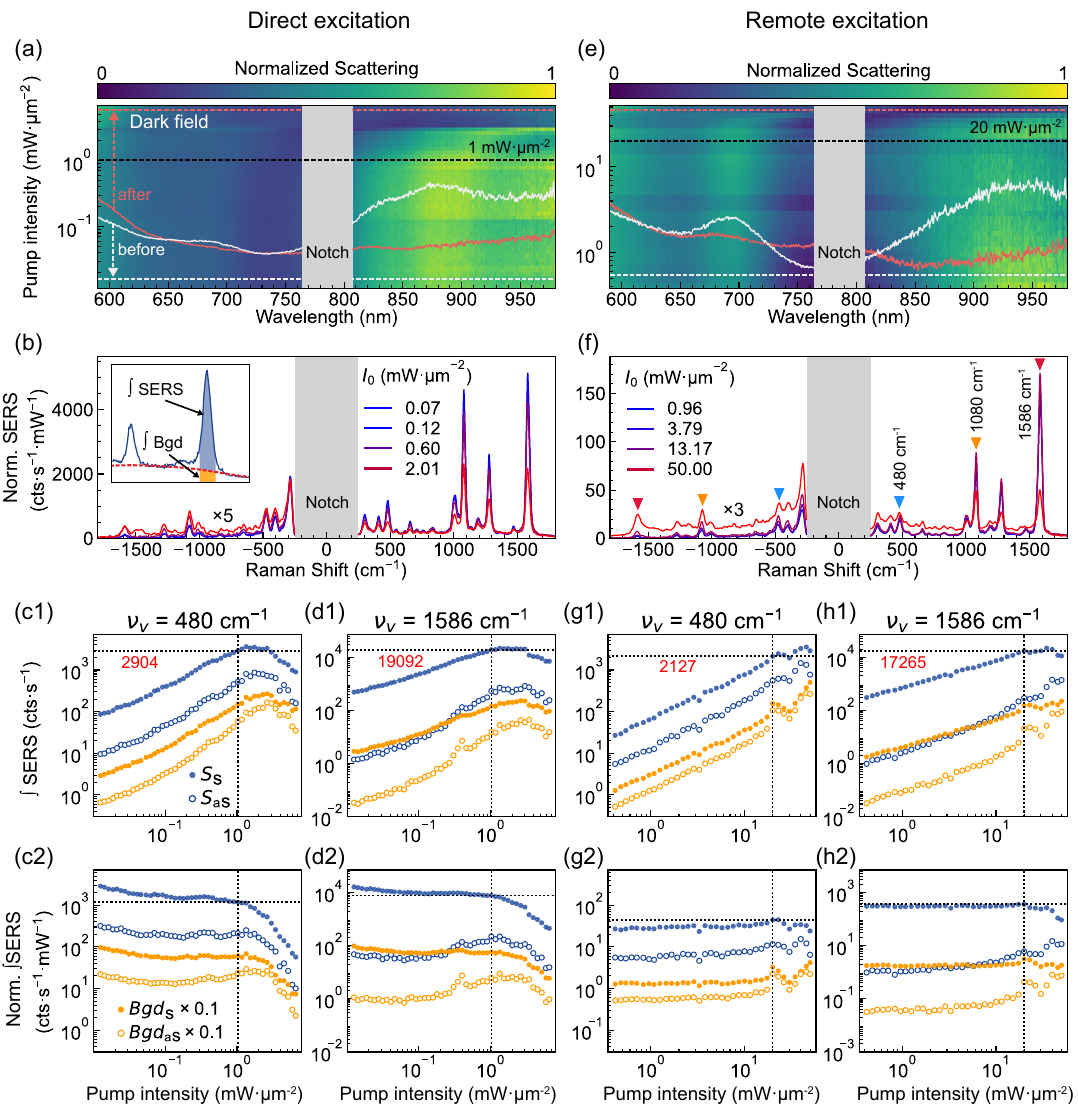}
  \caption{\textbf{Power dependence of SERS under direct and remote excitation.} Dark field (a, e) and SERS spectra (b, f) of representative BPT-functionalized NPoM nanogaps under (a, b) local and (e, f ) remote excitation. Two representative scattering spectra before (white) and after (red) the illumination are shown. Stokes and anti-Stokes spectra of the BPT molecules under direct (b) and remote (f) excitation of two representative NPoMs with different powers. 
  The inset in (b) shows the definition of Raman signal intensity (integrated area) and the exclusion of background photoluminescence. (c, d, g, h) The power-dependence of the (c1, d1, g1, h1) integrated and (c2, d2, g2, h2) normalized integrated SERS intensities of selected vibrational modes (480 and 1586 cm$^{-1}$) under direct (c, d) and remote (g, h) excitations. (c2, d2, g2, h2) The integrated SERS intensities were normalized by the pump power to reveal the signal scaling behavior. A flat curve (i.e., zero slope) indicates linear scaling and reflects stable SERS performance prior to damage. Blue filled (open) circles represent Stokes SERS (anti-Stokes SERS) signals. Orange filled (open) circles represent Stokes background signals (anti-Stokes background signals).}
  \label{fig:2}
\end{figure}

Although traditional studies on plasmonic systems have focused primarily on signal enhancement, achieving \textit{stable} maximum performance can be equally crucial for practical applications.
In Fig. \ref{fig:2}, we performed power-dependent measurements by gradually increasing the incident 785 nm laser power until discernible anomalies or irreversible reconstruction/damage emerged, as identified by their optical responses (i.e., dark-field and SERS spectra) under two different excitation schemes. 
We then recorded the maximum achievable SERS intensity prior to damage and defined the corresponding incident power as the optical damage threshold, from which the electric field limit  \( E_\mathrm{lim} \) could be inferred. We defined the \textit{damage} by observing a non-linear power law of the SERS intensity against the incident power.
The experimental protocol involved sequential dark-field scattering spectroscopy and SERS acquisition at each excitation power step, starting from 0.03~mW. This protocol was applied under both direct and remote excitation conditions, as demonstrated for representative single NPoMs under direct excitation (Figs.~\ref{fig:2}a--d) and remote excitation (Figs.~\ref{fig:2}e--h), respectively.
Crucially, to ensure a consistent total energy dose (Joules) across measurements and facilitate a fair comparison of damage thresholds, higher laser powers were applied with proportionally reduced exposure durations (details in Supplementary Fig. S8) \cite{jakobgiant2023}. 
Stokes (anti-Stokes) intensities for vibrational modes at 480 ($-480$) cm$^{-1}$ and 1586 ($-1586$) cm$^{-1}$ were quantified at each power level (Figs.~\ref{fig:2}c, d). After each SERS measurement, dark-field spectra were acquired to monitor whether the structural and plasmonic properties of the nanogaps remained intact.

\begin{figure}[!ht]
  \centering
  \includegraphics[width=0.8\textwidth]{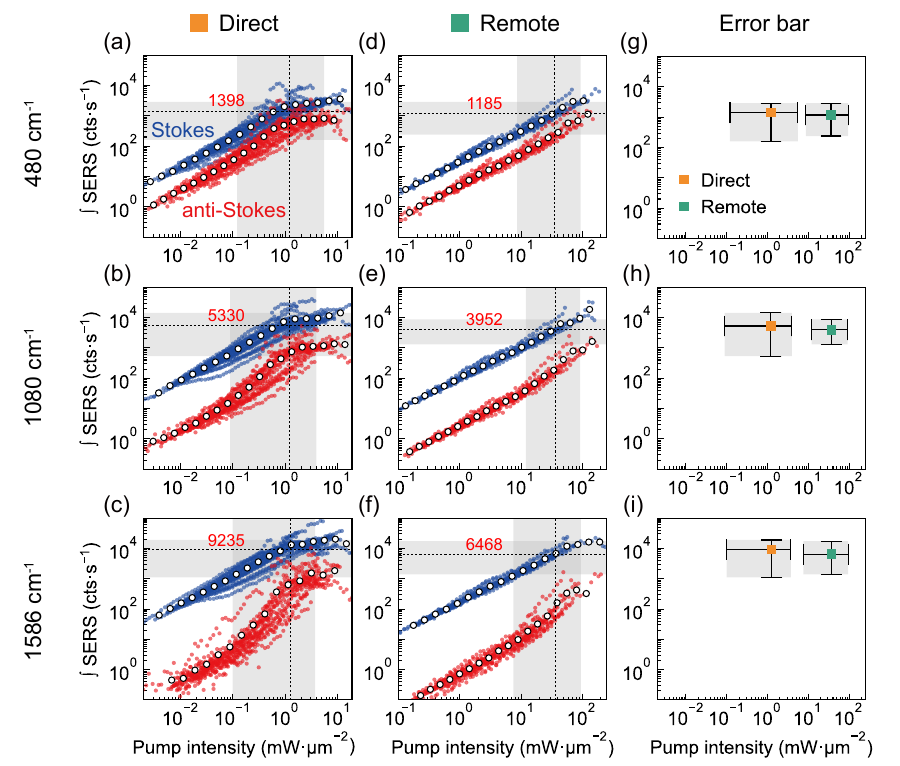}
  \caption{\textbf{Statistically probing photostability of SERS under different excitation schemes.} 
  (a–f) Power-dependent integrated SERS intensities of the Stokes (blue circles) and anti-Stokes (red circles) sidebands corresponding to the 480 cm$^{-1}$ (a, d), 1080 cm$^{-1}$ (b, e) and 1586 cm$^{-1}$ (c, f) vibrational modes under direct (a, b, c) and remote (d, e, f) excitation. Open circles represent the binned centers of mass of the corresponding data. (g–i) Distribution ranges of damage threshold powers and corresponding maximum SERS intensities for all nanoparticles under direct and remote excitation. Orange and green squares represent the average values under each excitation condition. 
  }
  \label{fig:3}
\end{figure}

Taking one single NPoM under direct excitation for instance  (Fig.~\ref{fig:2}a), we monitored its dark-field spectra under the increasing pump intensities. The full spectra appear discontinuous due to the notch filter employed in the experiment. The scattering response remained stable up to a pump intensity threshold of 1~mW$\cdot\mu \textrm{m}^{-2}$, beyond which the plasmon resonance near 850 nm began to redshift and eventually vanished, indicating the onset of optical damage. This spectral evolution is consistent with laser-induced nanogap morphology changes, such as facet growth~\cite{xomalis_controlling_2020,mertens_light_2017,sigle_monitoring_2015}, and thus serves as a sensitive indicator of structural damage. It has to be mentioned that, 1~mW$\cdot\mu \textrm{m}^{-2}$ is the threshold of this specific case of NPoM, the statistical analysis will be shown later in Fig. \ref{fig:3}.
More importantly, these stable dark-field scattering spectra were accompanied by stable SERS spectra (Fig.~\ref{fig:2}b). The SERS spectra were normalized by the input power, such that the overlapping curves indicate stable performance. For instance, we plotted three SERS spectra obtained under 1~mW$\cdot\mu \textrm{m}^{-2}$, along with another spectrum acquired above this threshold. The low-power SERS spectra exhibit nearly identical normalized intensities, whereas the above-threshold Stokes signal (i.e., 2.01~mW$\cdot\mu \textrm{m}^{-2}$ in this case) shows a remarkable intensity drop to approximately half. 
Other important evidences, such as increased background intensity, emergence of spurious peaks, and mode shifts, were shown and analyzed in Supplementary Fig. S4.

In Figs.~\ref{fig:2}c, d, we differentiated and integrated the SERS signals alongside the background contributions for two representative Raman peaks (480 and 1586~cm$^{-1}$; baseline removal shown in Fig.~\ref{fig:2}b inset, see Methods). The integrated SERS signals before (Figs.~\ref{fig:2}c1, d1) and after normalization (Figs.~\ref{fig:2}c2, d2) were analyzed as a function of pump intensity. As shown in Figs.~\ref{fig:2}c1, d1, both the integrated Raman signals (Stokes and anti-Stokes) and their backgrounds increase near-linearly with the pump intensity below the damage threshold (i.e., 1~mW$\cdot\mu \textrm{m}^{-2}$). Correspondingly, the normalized SERS intensities exhibit a nearly flat trend (Figs.~\ref{fig:2}c2, d2). Once above this threshold, the integrated SERS intensity drops sharply due to nanogap reconstruction, plateauing at approximately 2904 counts per second (cts s$^{-1}$) for the 480~cm$^{-1}$ peak, and 19092~cts s$^{-1}$ for the 1586~cm$^{-1}$ peak.
These absolute values represent the maximum achievable SERS in this system and provide important benchmarks for evaluating ultimate system performance. Furthermore, since SERS intensity scales linearly with excitation power (given the same geometry and identical molecules), these benchmarks are critical for fairly comparing different excitation schemes, which we would elaborate on later.


In contrast, we next applied the same measurement protocol to the nanogaps under remote excitation.
As shown by the scattering spectra in Fig.~\ref{fig:2}e, the remote excitation scheme requires a much higher pump intensity (around $20~\mathrm{mW} \cdot \mu \mathrm{m}^{-2}$) to induce structural reconstruction. This is due to the energy losses during SPP generation and propagation (see Supplementary Fig. S3).  
This further underscores the need to compare the maximum achievable SERS signals, as previously discussed, since NPoMs are randomly distributed at varying distances from the grating, and the power threshold for triggering structural instabilities can therefore differ significantly between individual particles. Similar to the direct-excitation case, the normalized Raman spectra with power above the threshold of  $20~\mathrm{mW} \cdot \mu \mathrm{m}^{-2}$ can be totally off from the weak-excitation ones (see the case with $50~\mathrm{mW} \cdot \mu \mathrm{m}^{-2}$ excitation in Fig. \ref{fig:2}f), manifesting as the geometric reconstruction.
A representative case of single NPoM nanogaps (Figs.~\ref{fig:2}g-h) similarly exhibited a stable regime characterized by a linear increase in SERS intensity up to approximately $20~\mathrm{mW} \cdot \mu \mathrm{m}^{-2}$ laser power density. The maximum achievable SERS intensity reached around 2127 cts s$^{-1}$ for the 480~cm$^{-1}$ peak and 17265 cts s$^{-1}$ for the 1586~cm$^{-1}$ peak.
Strikingly, this indicates that the SERS signal under remote excitation saturates at (and breaks after) nearly the same intensity limit as under local excitation. This similarity in maximum achievable SERS intensity suggests that the maximum electric field sustainable by the NPoM nanogap remains comparable regardless of the excitation scheme.
The much higher input power required for remote excitation results from the limited conversion efficiency and propagation losses of SPPs.


Plasmonic nanogaps inherently suffer from varying performances from case to case due to the local fine morphologies, e.g., roughness, corners, protrusions, local environment, etc. A statistical study of the thresholds and maximum SERS intensities of numerous NPoM can be relevant in this context. To do so,  in  Fig. \ref{fig:3}, we standardized \cite{jakobgiant2023} the data by normalizing the SERS intensity of different nanoparticles based on their Raman signal strength at the lowest excitation power, thereby allowing the aggregation of their power-dependent SERS responses (see Supplementary S6 for details).
Consistent with the representative cases in Fig.~\ref{fig:2}, our statistical analysis of 38 nanoparticles in Fig.~\ref{fig:3} shows that the maximum attainable SERS intensities under both direct and remote excitation schemes plateau at similar levels across three distinct Raman modes, i.e., 480, 1080, and 1586 cm$^{-1}$. Specifically, the maximum SERS intensities at 480, 1080, and 1586 cm$^{-1}$ reached 1398 cts s$^{-1}$ (1185 cts s$^{-1}$), 5330 cts s$^{-1}$ (3952 cts s$^{-1}$), and 9235 cts s$^{-1}$ (6468 cts s$^{-1}$) for direct (remote) excitation, respectively. Additionally, power-normalized data for Fig.~\ref{fig:3} are provided in Supplementary Fig. S5.

The error bars of Figs.~\ref{fig:3}g--i represent the damage thresholds and maximum achievable SERS intensity of every single NPoM. 
They suggest that despite distinct excitation pathways, the maximum electric field sustainable within the nanogap remains indeed comparable between schemes. 
Remote excitation requires higher input power due to coupling and propagating losses, but both schemes converge to similar SERS saturation levels, confirming the field-driven nature of the limitation. This could challenge the intuition that remote excitation may enhance the maximum SERS output or raise the laser damage threshold via reduced thermal effects. 

\begin{figure}[!ht]
  \centering
  \includegraphics[width=0.55\textwidth]{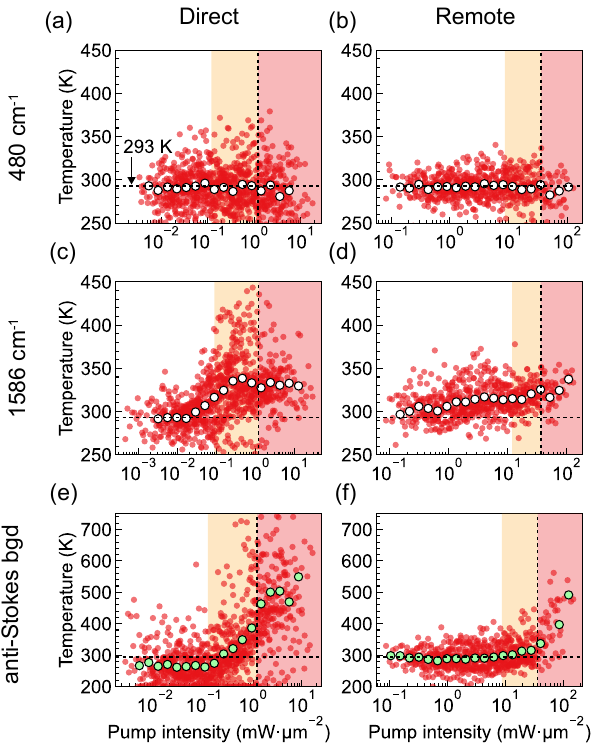}
  \caption{\textbf{Probing the local temperature within the NPoM.} Local temperatures extracted using two methods: the molecular Stokes/anti-Stokes intensity ratio (a–d) and the anti-Stokes background fitting from gold (e, f). Panels (a, c, e) correspond to direct excitation conditions, while (b, d, f) correspond to remote excitation. White and green circles represent the binned centers of mass, while red dots are the data points. }
  \label{fig:4}
\end{figure}

While previous studies have often attributed nanogap damage to laser-induced heating or field-driven morphological changes, the precise origin of the damage threshold remains debated.
To quantify the thermal effects in the power-dependent analysis discussed above, we non-invasively extract the local temperature by analyzing the ratio of anti-Stokes to Stokes Raman intensities based on the Boltzmann distribution in Fig.~\ref{fig:4}. The intensity ratio reflects the population of vibrational energy states, which depends on temperature as follows \cite{skinner_all--one_2025}:

\begin{equation}
\label{eq:boltzmann}
\frac{I_{\text{aS}}}{I_{\text{S}}} = A_i \left( \frac{\nu_i + \nu_v}{\nu_i - \nu_v} \right)^3 \exp\left( -\frac{h \nu_v}{k_B T} \right)
\end{equation}

\noindent Here, $A_i$ is an asymmetry factor arising from the combined effects of resonance effects and the optical response of the Raman detection system (see Methods). \(I_{aS}\) and \(I_S\) represent the anti-Stokes and Stokes Raman intensities, respectively; \(\nu_i\) is the excitation laser frequency; \(\nu_v\) is the vibrational mode frequency; \(h\) is Planck’s constant; \(k_B\) is Boltzmann’s constant; and \(T\) is the absolute temperature. Since the anti-Stokes intensity increases with temperature due to higher vibrational state populations, measuring this ratio allow for direct calculation of the local sample temperature.


As shown in Fig.~\ref{fig:4}, temperature estimates based on, for example, 480 and 1586 cm$^{-1}$ modes, reveal distinct excitation-dependent trends. For the lower-frequency mode (480 cm$^{-1}$), nanogap temperatures under both excitation conditions remained near room temperature (293 K) across the full varying power range, suggesting negligible heating. However, for the higher-energy 1586 cm$^{-1}$ mode, direct excitation led to a steady temperature increase up to 350 K near the damage threshold, whereas remote excitation induced a more modest rise to 320 K. 

These results were independently revealed via electronic Raman background analysis (Figs.~\ref{fig:4}e, f), which refers to the Eq. 2 \cite{jollanseffective2020,baffouAntiStokesThermometryNanoplasmonics2021b}:

\begin{equation}
\label{eq:background}
I_{\mathrm{aS}}^{\mathrm{bgd}}(\Delta \nu) = B \exp\left( \frac{-h \Delta \nu}{k_{\mathrm{B}} T_e} \right),
\end{equation}
\noindent where  \( I_{\mathrm{aS}}^{\mathrm{bgd}}(\Delta \nu) \) represents the anti-Stokes background intensity at the Raman shift \(\Delta \nu\), typically measured in counts or intensity units; \(\Delta \nu\) is the Raman shift in units of cm\(^{-1}\), corresponding to the frequency offset of the scattered light relative to the excitation source; \( B \) is a proportionality constant reflecting the enhancement and detection efficiency of the emission process; and \( T_e \) is the extracted electronic temperature, representing the effective temperature of the electrons in the metallic nanostructure (see Methods).

Figs.~\ref{fig:4}e, f show that before damage, direct excitation raised temperatures in the nanogap beyond 450~K (an increase of 160~K), while remote excitation caused a much smaller rise (30~K), confirming that remote excitation suppresses laser heating, aligning with the previous experiment~\cite{evans_quantifying_2017}. 
The comparable absolute SERS intensities (Fig. \ref{fig:3}), despite reduced heating in the remote case, suggest that the critical threshold remains unchanged. 
Since lower heating (Fig. \ref{fig:4}) does not raise the damage threshold, equilibrium thermal effects are unlikely to dominate the damage mechanism.  This evidence reinforces that the structural degradation is driven by non-thermal, field-induced processes such as plasmon-driven atomic rearrangement or interfacial modification within the nanocavity.

\begin{figure}[!ht]
  \centering
  \includegraphics[width=0.6\textwidth]{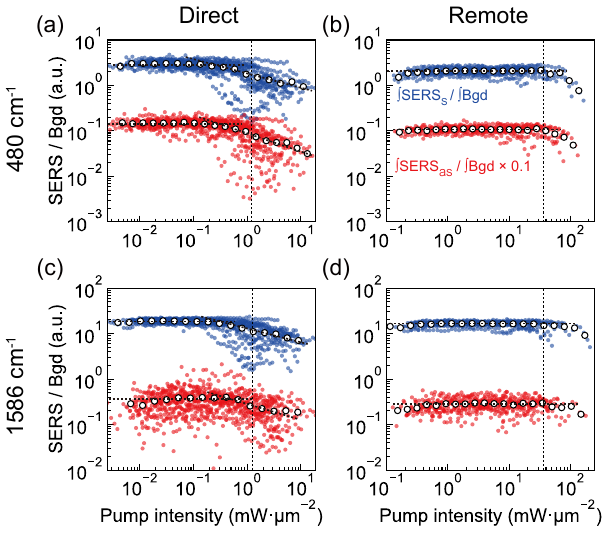}
  \caption{\textbf{Signal-to-noise ratio of the Raman signal.} Power-dependent SNR under the direct (a, c) and remote (b, d) excitation schemes. SNR were calculated with the Raman peaks with the shifts of 480 cm$^{-1}$ (a, b) and 1586 cm$^{-1}$ (c, d). White dots are the binned center of mass, and the dots are the data points.}
  \label{fig:5}
\end{figure}

Finally, we quantified the SNR of SERS under both excitation schemes in Fig.~\ref{fig:5}. The SNR is a critical metric that determines the quality and fidelity of the measured signal. It can be defined as the ratio between the Raman peak intensity and the underlying spectral background (see insets of Fig.~\ref{fig:2}b). We statistically analyzed the SNR of two Raman peaks (480 and 1586~cm$^{-1}$). In the low-intensity range, both excitation schemes have very similar SNR, as indicated by the centers of mass of each plot.
When the pump intensities are increased, the structural damages in both cases are consistently accompanied by a sharp drop in SNR, rendering degraded nanogaps unsuitable for reliable sensing. 
More strikingly, the SNR under direct excitation began to deteriorate even before the damage threshold, whereas remote excitation preserved a stable SNR over the whole range until crossing the threshold. This can be correlated to the more prominent temperature increase in the local excitation than in the remote case.
Near the breakdown threshold of each excitation scheme, the SNR under remote excitation remained at 2.06 (15.05), while direct excitation dropped to 1.63 (11.16), representing a 26\% (34\%) improvement for the Raman peak at 480 (1586) cm$^{-1}$ (See Fig. S9 in SI for detailed analysis).
These results highlight that remote excitation offers improved signal fidelity and reduced variability, while still achieving comparable maximal SERS intensity. Although it does not raise the damage threshold, these advantages make it particularly compelling near power limits, where inherently weak SERS signals usually demand high-intensity excitations and the enhanced SNR of remote excitation can notably improve the performance.



In the experiments, we primarily studied 150 nm Au NPoMs, as these structures support stronger radiative plasmon modes and produce higher SERS signals—enabling higher-SNR measurements. In contrast, other structures with different sizes, shapes, or even materials may exhibit very different SERS behavior due to the varying resonances and different modal properties.\cite{baumberg_extreme_2019,liBoostingLightMatterInteractions2024a} This raises a fundamental question: do both excitation schemes generally support the same maximum local field limit, \( E_\mathrm{lim} \), and therefore produce similar maximal SERS output?
To explore this, as an example, we performed power-dependent measurements on 80 nm NPoMs under direct excitation scheme (Supplementary Fig.~S7). The results show that their maximal SERS intensity remains substantially lower than that of 150 nm NPoMs, indicating a much lower supportable field limit $E_{\rm lim}$. Additionally, it asks for higher input powers to reach the $E_{\rm lim}$ due to the lower field enhancement. This indicates that the maximum achievable SERS is determined by the structural stability of the material system against optical and thermal effects. 
In particular, weaker enhancement does not imply a higher damage threshold, and the supportable local field should vary across systems depending on material, surface chemistry, and geometries. This work generally proves the existence of a general $E_{\rm lim}$ for each specific nanoplasmonic system.
In this regard, seeking the most robust and outstanding SERS substrates with maxima achievable SERS intensities, systematic extraction and quantification of $E_\mathrm{lim}$ across different nanostructures and excitation wavelengths warrants further investigation. \cite{hu2025alchemically}

In addition, although plasmonic systems with strong enhancement have been considered as promising candidates for lowering energy consumption in nonlinear optical applications \cite{kauranen_nonlinear_2012}, nanogap systems often fail to reach their full potential due to instability under intense femtosecond laser excitation.
One might believe that heating effects and geometric reconstruction played important roles, and remote excitation schemes with reduced local heating were believed to alleviate these issues \cite{li_duplicating_2020,evans_quantifying_2017}. However, current findings challenge this intuitive belief, suggesting that the local electric field—rather than temperature—is the primary factor governing structural and optical stability. Given the fixed maximum local field \( E_{\rm lim} \) that a nanogap can sustain, advancing nonlinear plasmonics will require either the discovery of new physical mechanisms~\cite{de_luca_free_2021,hu_low-power_2024} or the development of novel materials with enhanced nonlinear susceptibilities \( \chi^{(n)} \)~\cite{jelver_nonlinear_2023,cox_electrically_2014}.

In conclusion, our statistical benchmarking shows that both remote and direct excitation yield comparable maximal SERS intensities, revealing a general sustainable electric field limit that plasmonic nanogaps must obey, independent of excitation scheme or heating differences.
However, remote excitation offers improved SNR, facilitating the reliable readout of weak signals in complex environments.
These results challenges some of the prevailing assumptions that remote excitation enhances photostability via reduced heating and underscores the need for new materials or nanogap designs capable of sustaining stronger fields or optically more active. Our findings provide key design guidelines for advancing plasmonic sensing and nonlinear nanophotonics.

\section*{Methods}

\subsection*{Optical measurements}

\textbf{Dark-field scattering measurements.}
Dark-field scattering spectroscopy and imaging were performed using home-built microscope system with a commercial illuminator (BX53, Olympus). White light from a halogen lamp was directed through a dark-field module to form annular illumination, which was then focused onto the sample through a 100× dark-field objective lens (Olympus, NA = 0.9) for dark-field excitation. The scattered light was collected by the same objective and subsequently passed through a beam splitter (Chroma), where 10 percent of the light was directed to a CMOS camera for dark-field imaging and 90 percent was coupled into a confocal spectrometer (HORIBA iHR320) for spectral acquisition using an integration time of 1 s. All acquired dark-field spectra were normalized by dividing with the reference light source spectrum.

\textbf{Direct and remote SERS measurement.}
For SERS measurement we employed the same microscope system as in the dark-field measurement, except an additional beam splitter used for laser excitation, and two 785 nm notch filters (Thorlabs) added in the collection path for the laser blocking. For direct SERS excitation, both the excitation spot and collection area were aligned on the target nanoparticles; for remote SERS measurements, the excitation was focused on the grating while maintaining the collection area on the nanoparticles.



During power-dependent Raman measurements, the laser power was controlled using a motorized optical attenuator. Detailed optical setup is shown in Supplementary Fig. S2.


\subsection*{SERS baseline removal}
For quantitative spectral analysis, an Asymmetric Least Squares (AsLS) background fitting algorithm was employed to automatically separate the Raman signal from the background. The integrated intensity of the Raman peak was calculated within a $\pm 35~\mathrm{cm}^{-1}$ (1.5~nm) window around the peak center. The background signal was integrated using the same window.

\subsection*{Temperature extraction}
To evaluate the system’s thermal state, the molecular temperature within the nanocavity was extracted from the anti-Stokes to Stokes intensity ratio using a Boltzmann model (Eq.~\eqref{eq:boltzmann}) incorporating a calibration factor \( A_i \), which accounts for nanocavity resonance and the optical system’s response. After subtracting the background, peak integrals were employed to calculate the intensity ratios. Assuming the initial local temperature corresponds to room temperature, the calibration factor \( A_i \) for each vibrational mode was determined from spectra measured at low excitation power. These factors were subsequently applied to estimate the local temperature rise under higher excitation powers, provided the nanocavity remained intact.
To evaluate the local electronic temperature $T_e$ of gold nanostructures, the broadband anti-Stokes background emission was fitted using a Boltzmann-type model as described in (Eq.~\eqref{eq:background}). The proportionality factor $B$ reflects the enhancement and detection efficiency of the emission process. In the logarithmic scale, the intercept of the fitted curve is determined by $B$, while the slope corresponds to the electronic temperature $T_e$. This approach is particularly advantageous under conditions where the molecular Raman peaks are weak or unresolved, such as at high excitation powers or during molecular degradation. It assumes that the electronic and lattice temperatures are in thermal equilibrium, which is typically valid under quasi-steady-state illumination. 

\subsection*{Numerical simulation}
The scattering and the field enhancement of the 150 nm NPoM (with 10 nm bottom facet) were calculated with COMSOL Multiphysics. The refractive index of the BPT molecule is 1.4. A 2 nm-thick CTAC molecular layer (refractive index $=$ 1.5) was assumed to coat the 150 nm nanosphere. Permittivity of gold was taken from Olmon et.al.\cite{olmon2012optical} P-polarized, oblique incident light with an angle of 80 deg was considered.

\begin{acknowledgement}
    
This work was supported by the National Key Research and Development Program of China (Grant No. 2024YFA1409900) and the National Natural Science Foundation of China (Grant Nos. 62475071 and 52488301).

\end{acknowledgement}

\begin{suppinfo}

\noindent
Sample Preparation and Surface Morphology; Optical Setup; Statistics of Remote Excitation Transmission Efficiency; Power-Dependent Raman Data of Additional Individual NPoM; Power Dependence of Power-Normalized SERS Intensity; Calibration of Effective In-Coupling Power in NPoMs; Power-Dependent Measurements of 80-nm NPoMs;  Exposure settings for direct and remote excitation; Average Signal-to-Noise Ratio Near Threshold.

\end{suppinfo}

\bibliography{remote_hetao}

\end{document}